\documentclass[aps,pra,twocolumn,floatfix]{revtex4-1}
\usepackage{graphicx}
\usepackage{epstopdf}
\usepackage{amssymb,amsmath}
\usepackage{bm}
\begin{document}
\title{Electromagnetically Induced Transparency in strongly interacting Rydberg Gases}

\author{C. Ates}
\affiliation{Max Planck Institute for the Physics of Complex Systems, N\"othnitzer Strasse 38, 01187 Dresden, Germany}
\author{S. Sevin\c{c}li}
\affiliation{Max Planck Institute for the Physics of Complex Systems, N\"othnitzer Strasse 38, 01187 Dresden, Germany}
\author{T. Pohl}
\affiliation{Max Planck Institute for the Physics of Complex Systems, N\"othnitzer Strasse 38, 01187 Dresden, Germany}

\date{\today}

\begin{abstract}
We develop an efficient Monte-Carlo approach to describe the optical response of cold three-level atoms in the presence of EIT and strong atomic interactions. In particular, we consider a "Rydberg-EIT medium" where one involved level is subject to large shifts due to strong van der Waals interactions with surrounding Rydberg atoms. We find excellent agreement with much more involved quantum calculations and demonstrate its applicability over a wide range of densities and interaction strengths. The calculations show that the nonlinear absorption due to Rydberg-Rydberg atom interactions exhibits universal behavior.
\end{abstract}

\maketitle
The effect of electromagnetically induced transparency (EIT) \cite{EIT_rev} in light-driven multi-level systems continues to play a pivotal role in quantum and nonlinear optics. Enabling slow light propagation and thus long photon interaction times at low loss levels \cite{luim00,hafi99}, EIT media provide a promising route to applications in optical communication and quantum information science. Optical nonlinearities, however, typically arise from higher order light-atom interactions, such that realizations of such applications at very low light intensities \cite{bhb09} remain challenging. 
In this respect, recent experimental studies of EIT in cold Rydberg gases \cite{moja07,moba08,wepr08,rahe09,pgw10} are opening up new perspectives for nonlinear optics on a few photon level. Exploiting the exaggerated properties of Rydberg atoms and, in particular, the strong interactions among the atoms, nonlinear phenomena can be greatly enhanced in ultracold Rydberg gases. 

Recently, several different methods have been used to study laser-driven interacting gases \cite{fde10, sgh10,welo08,pode10,slm10}. A simultaneous treatment of EIT and long-range interactions, however, poses additional challenges. Common meanfield approaches, as successfully applied to radiation trapping effects on EIT \cite{fde10}, are found to fail \cite{sgh10} due to the large strength and enormous range of the van der Waals interaction, which lead to non-negligible correlations between the atoms. On the other hand, exact descriptions, based on Hilbert space truncation by interaction-blocked many-body states \cite{welo08,pode10}, are also inapplicable since there always remains an exponentially large number of many-body states involved in the interaction-free probe transition.

Here we present a theoretical approach that allows to obtain the fully correlated steady state populations via classical Monte Carlo sampling. This is shown to yield the nonlinear optical response to classical light fields in the presence of arbitrarily strong atomic interactions. A comparison to reduced density matrix calculations shows very good agreement for small and moderate densities. Upon proper scaling the simulation results reveal a universal behavior of the nonlinear absorption, which illustrates the role of Rydberg atom interactions in the emergence of dissipative photon-photon interactions. 

\begin{figure}[t]
\begin{center}
\resizebox{0.9\columnwidth}{!}{\includegraphics{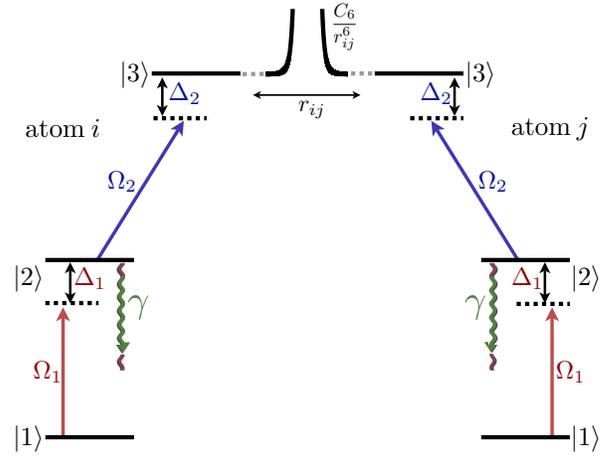}}
\caption{Illustration of the considered three-level ladder scheme. Two laser fields successively couple the states $|1\rangle$, $|2\rangle$ and $|3\rangle$. For isolated  atoms, EIT is realized on two-photon resonance $\Delta_1=-\Delta_2$. The van der Waals interaction shift $C_6/r_{ij}^6$ between atoms in the Rydberg state $|3\rangle$ modifies this ideal transparency and leads to a nonlinear optical response of the medium.  \label{fig1}}
\end{center}
\end{figure}

The considered Rydberg-EIT level scheme  is shown in Fig.\ref{fig1}. 
The signal laser couples the ground state $|1\rangle$ to a low lying state $|2\rangle$ with Rabi frequency $\Omega_1$. State $|2\rangle$ is coupled by a  strong control laser ($\Omega_2 >\Omega_1$) to a highly excited Rydberg state $|3 \rangle$. The resulting dynamics of a gas composed of $N$ such independent atoms is, thus, governed by the Hamiltonian
\begin{eqnarray}
\label{eq1}
H_0 &=& -\frac{1}{2}\sum_{i=1}^N ( \Delta_1 | 2_i \rangle \langle 2_i | + (\Delta_1+ \Delta_2) | 3_i \rangle \langle 3_i | \nonumber \\ 
&&- \Omega_1 | 2_i \rangle \langle 1_i | - \Omega_2 | 3_i \rangle \langle 2_i | + \text{h.c.} )\;.
\end{eqnarray}
In addition, the intermediate state $|2\rangle$ radiatively decays with a rate $\gamma$, whereas spontaneous decay of the long-lived Rydberg level can safely be neglected. 
For resonant driving, each atom settles into a dark state, $|d_i\rangle\sim\Omega_2|1_i\rangle-\Omega_1|3_i\rangle$, which is immune to the laser coupling and radiative decay \cite{EIT_rev}. Consequently, the complex susceptibility $\chi_{12}$ of the lower transition vanishes and the medium becomes transparent. 

In the presence of van der Waals interactions 
\begin{equation}
\label{eq2}
U = \sum_{i<j}  \frac{C_6}{|\mathbf{r}_i -\mathbf{r}_j|^6} | 3_i 3_j \rangle \langle 3_i 3_j |
\end{equation} 
between atoms in state $|3\rangle$ the situation becomes more complex. Note that the van der Waals coefficient $C_6\sim n^{11}$ strongly increases with the atom's principal quantum number $n$. For typical values $n\sim 60$ this yields an enhancement of about $10^{11}$ compared to interactions in low-lying states, implying a drastically different excitation dynamics.
On the one hand, the resulting level shifts lead to a strong suppression of Rydberg excitation\cite{Rydexp}, due to an excitation blockade of close atoms \cite{lfc01}. On the other hand, the interactions perturb the atomic dark states \cite{mmm08,mlw09}, and admix the dissipative intermediate state \cite{pgw10,sgh10}. To account for these effects, we start from the von Neumann equation
\begin{equation}
\label{eq3}
i \dot{\rho}^{(N)} =  \left[ H_0 + U \right] -i \mathcal{L}[\rho^{(N)}] \; ,
\end{equation}
for the $N$-body density matrix $\hat{\rho}^{(N)}$ of the gas.
The Lindblad operator $\mathcal{L}$ describes spontaneous decay of the intermediate state, but can also include finite laser band widths, denoted by $\gamma_{12}$ and $\gamma_{23}$, respectively.  

To obtain the steady state populations of eq.(\ref{eq3}), we transform it to a many-body rate equation.  
For clarity we start from the simple case $C_6=0$ for which the $N$-particle density matrix factorizes, and eq.(\ref{eq3}) reduces to a set of single-atom optical Bloch equations. Upon adiabatic elimination the dynamics of the atomic level populations follows 
\begin{equation}
\label{eq4}
\frac{\text{d}}{\text{d}t} \left(
\begin{array}{c}
\rho_{11}^{(i)} \\ \rho_{22}^{(i)} \\ \rho_{33}^{(i)}
\end{array}
\right) =
\left(
\begin{array}{ccc}
-a_{11} & a_{12} & a_{13} \\
a_{21} & -a_{22} & a_{23} \\
a_{31} & a_{32} & -a_{33}
\end{array}
\right)
\left(
\begin{array}{c}
\rho_{11}^{(i)} \\ \rho_{22}^{(i)} \\ \rho_{33}^{(i)}
\end{array}
\right) \, ,
\end{equation}
where the coefficients $a_{\alpha \beta}$ are straightforwardly obtained as a function of the laser parameters. 

This is a common approximation to the long-time dynamics of two-level systems \cite{ae75}. Applications of the described elimination procedure to three-level atoms are, however, rather scarce, since it often leads to negative transition rates \cite{wila60}. To resolve this obstacle we propose a linear transformation that removes the negativity and, at the same time, preserves the correct steady states of the underlying von Neumann equation (\ref{eq3}).  
For $\Omega_1<\Omega_2$ this is accomplished by adding a correction matrix 
\begin{widetext}
\begin{equation}\label{trafo}
\Delta \mathbf{a} = \left(
\begin{array}{ccc}
0 & \sigma_{21}(1 - R_{21}) + \sigma_{32} & \sigma_{31} (1 - R_{31} )\\
-\sigma_{21} + \sigma_{31} - \sigma _{32} {}R_{32} & 
  -\sigma_{21}( 1- R_{21} )- \sigma_{31} R_{31} &
   \sigma_{21} R_{21} +  \sigma_{32} (1- R_{32}) \\
\sigma_{21} - \sigma_{31} + \sigma_{32} R_{32} & -\sigma_{32} + \sigma_{31} R_{31} & 
   -\sigma_{21} R_{21} -\sigma_{31}(1 - R_{31})  -\sigma_{32} (1- R_{32})
\end{array}
\right) \; ,
\end{equation}
\end{widetext}
to the original coefficient matrix ${\bf a}$ in eq.(\ref{eq4}), where $\sigma_{\alpha \beta} = (|a_{\alpha \beta}| - a_{\alpha \beta})/2$ and $R_{\alpha \beta}$ are free parameters. With this definition of the $\sigma_{\alpha \beta}$ any negative rate coefficient $a_{\alpha\beta}$ is set to zero, while the additional terms compensate for the according changes of the steady states.  Consequently, the parameters $R_{\alpha \beta}$ are chosen such that the transformed rate equations $\dot{\bm \rho}^{(i)}=({\bf a}+\Delta{\bf a})\bm{\rho}^{(i)}$ yield steady states identical to those of the original equation $\dot{\bm \rho}^{(i)}={\bf a}\bm{\rho}^{(i)}$. Explicitely, this condition gives
\begin{subequations}
\begin{eqnarray}
R_{21} &=& \frac{a_{33} \left(a_{21} + a_{22} \right) + a_{23} \left( a_{31} - a_{32} \right)}{a_{31} \left( a_{22} +a_{23}  \right) + a_{21} \left( a_{32} + a_{33} \right)} \\
R_{31} &=&  \frac{a_{22} \left(a_{31} + a_{33} \right) + a_{32} \left( a_{21} - a_{23} \right)}{a_{31} \left( a_{22} +a_{23}  \right) + a_{21} \left( a_{32} + a_{33} \right)} \\
R_{32} &=&  \frac{a_{21} \left(a_{32} + a_{33} \right) + a_{31} \left( a_{23} + a_{33} \right)}{a_{22} \left( a_{31} +a_{33}  \right) + a_{32} \left( a_{21} - a_{23} \right)} 
\end{eqnarray}
\end{subequations}
for the transformation coefficients in eq.(\ref{trafo}).

Having established a proper rate equation description for the single particle dynamics, the approach can be extended to interacting atoms.
As shown in \cite{app07}, this is straightforwardly accomplished by replacing the upper detuning of each atom by $\Delta_2^{(i)}=\Delta_2-\sum^{\prime}_{j\neq i}U_{ij}$. Here, the sum only runs over atoms in the Rydberg state $|3\rangle$, such that the local detunings and, hence, the individual atomic transition rates become dynamical variables that depend on the entire many-body state of the system. As a result one obtains a many-body rate equation where the dynamics of the $i$th atom is governed by its individual transition rates $\tilde{a}_{\alpha\beta}^{(i)}$ which depend on the actual Rydberg atom configuration through the local detuning $\Delta_2^{(i)}$. Although this rate equation still covers the exponentially large number of all $3^{N}$ many-body states, it can be efficiently solved via classical Monte-Carlo sampling. Starting from the initial state with all atoms in their ground state $|\alpha\rangle=|1\rangle$, the steady state is obtained by performing repeated random transitions according to the probabilities $p_{\alpha \to \beta}^{(i)}=\delta t\cdot \tilde{a}_{\alpha\beta}^{(i)}$ to make a transition within a time step $\delta t$.
In this way, we are able to obtain the fully correlated $N$-body state-distribution of the particles, and in particular the average level populations $\rho_{\alpha\alpha}=N^{-1}\sum_i\rho_{\alpha\alpha}^{(i)}$. 
Since
\begin{equation}
\label{ab_pop}
\text{Im} (\rho_{12}) = \frac{\gamma}{\Omega_1} \rho_{22} \; ,
\end{equation} 
the Monte-Carlo approach also allows to determine the imaginary part of the complex optical susceptibility 
\begin{equation}
\label{suscept}
\chi_{12} = \frac{2 \mu_{12}^2 \rho_0}{\epsilon_0 \hbar \Omega_1} \rho_{12} \, ,
\end{equation}
where $\mu_{12}$ denotes the dipole moment of the probe transition and $\epsilon_0$ is the permittivity of vacuum. In addition, the resonant real part of $\chi_{12}$ can be obtained from the nonlinear absorption spectrum using the Kramers-Kronig relations.

\begin{figure}
\begin{center}
\includegraphics[width=0.95\columnwidth]{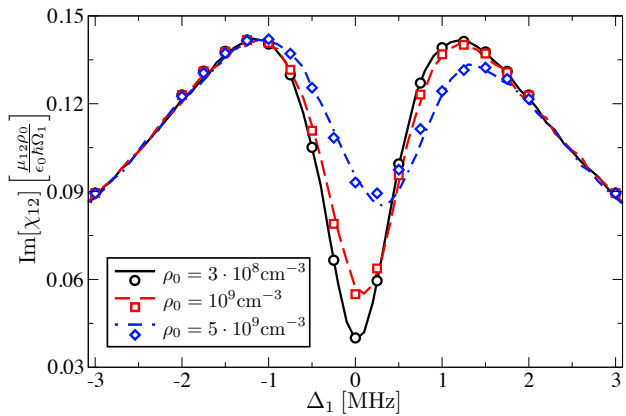}
\caption{Calculated probe beam absorption spectra for a Rubidium Rydberg-EIT medium at various densities. The lines show results of the Monte-Carlo simulations compared to quantum calculations based on a reduced-density matrix expansion \cite{sgh10} (symbols). The atoms are resonantly ($\Delta_2 =0$) excited to $55S$ Rydberg states. The probe and coupling beams have Rabi frequencies of $\Omega_1=1$MHz and $\Omega_2=2$MHz, respectively, with linewidths of $\gamma_{12}= \gamma_{23}=100$kHz.} 
\label{fig2}
\end{center}
\end{figure}

In the following we consider the specific case of a Rubidium Rydberg gas. The atoms are excited by the probe laser ($\Omega_1$) from the $|5S_{1/2}\rangle=|1\rangle$ ground state to the $|5P_{3/2}\rangle=|2\rangle$ intermediate state, while the coupling laser resonantly ($\Delta_2=0$) drives the transition between $|5P\rangle$ and a $|nS_{1/2}\rangle=|3\rangle$ Rydberg state with $\Omega_2 = 2$MHz. We include the intermediate state decay of $\gamma = 6.1$MHz and further assume realistic linewidths \cite{pgw10} of $\gamma_{12} = \gamma_{23} = 100$kHz for both beams.

Fig.\ref{fig2} shows the calculated absorption spectrum at different densities for $\Omega_1=1$MHz and $n=55$. In order to check our results at low densities we have also performed simulations based on a reduced-density expansion of the von Neumann equation (\ref{eq3}), as described in \cite{sgh10}. 
The excellent agreement between these entirely different calculations attests to the quality of both approaches. At the lowest density, Rydberg-Rydberg atom interactions are ineffective, giving a small resonant absorption due the finite laser linewidths $\gamma_{12}$ and $\gamma_{23}$. At higher densities the interactions lead to a significant suppression of the resonant transmission. In accord with recent experiments \cite{pgw10}, the position and width of the absorption minimum are, however, largely unaffected by the interactions, which are as large as $C_6(4\pi\rho/3)^2\approx20$MHz at the highest density of $5\cdot10^9$cm$^{-3}$. Apparently, this is due to the van der Waals blockade that prohibits simultaneous Rydberg excitation of close atoms.

\begin{figure}[b!]
\begin{center}
\includegraphics[width=0.95\columnwidth]{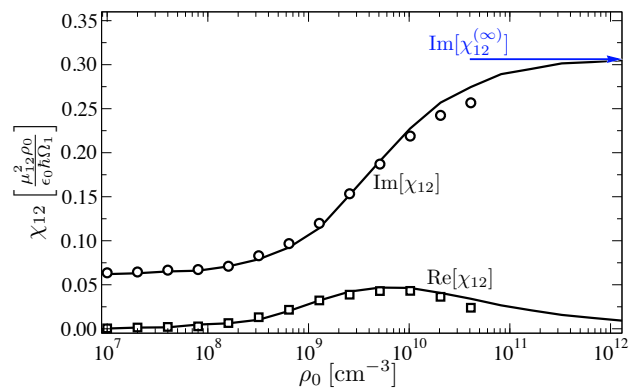}
\caption{Complex susceptibility $\chi_{12}$ as a function of the gas density $\rho_0$ under EIT conditions, $\Delta_1=\Delta_2=0$. The reduced density matrix calculations (symbols) start to deviate from the Monte Carlo results (lines) for densities $\gtrsim 10^{10}$cm$^{-3}$. Remaining parameters are identical to those of Fig.\ref{fig2}.}
 \label{fig3}
\end{center}
\end{figure}

The density dependence of the complex susceptibility is shown in Fig.\ref{fig3}. For low and moderate densities our Monte Carlo calculations and the reduced-density matrix expansion \cite{sgh10} give consistent results for both the real and imaginary part of $\rho_{12}$ (and hence of $\chi_{12}$). As expected from a low-density expansion the latter tends to deviate with increasing densities, with significant deviations occurring only at rather high densities $\gtrsim10^{10}$cm$^{-3}$. As we show in the following, the Monte-Carlo calculations yield the expected high density limit, and should thus applicable for arbitrary densities.

As the density increases the imaginary part of $\chi_{12}$ starts to saturate while the real part develops a maximum and goes to zero at high densities.
In fact, this behavior can be understood from simple arguments. At very high densities, a single Rydberg atom strongly shifts a large number of surrounding atoms out of resonance. As a consequence the Rydberg laser appears far detuned for the majority of atoms, such that they act as an effective two-level medium. Taking the limit $\Delta_2^{(i)}\rightarrow\infty$ the asymptotic high-density steady state value of $\rho_{12}$, hence, approaches
\begin{eqnarray}
\text{Im}[\rho_{12}^{(\infty)}] &=& \frac{\Omega_1 \gamma}{g_{12}\gamma^2 + g_{23} \Omega_1^2} 
\label{scale_chi}
\end{eqnarray}
where $g_{12} = (\gamma + \gamma_{12})/\gamma$, $g_{23} = (2 \gamma +3 \gamma_{23})/(\gamma + \gamma_{23})$. This simple  limit is indicated by the horizontal arrow in Fig.\ref{fig3} and gives very good agreement with the Monte Carlo simulations.

\begin{figure}
\begin{center}
\includegraphics[width=0.99\columnwidth]{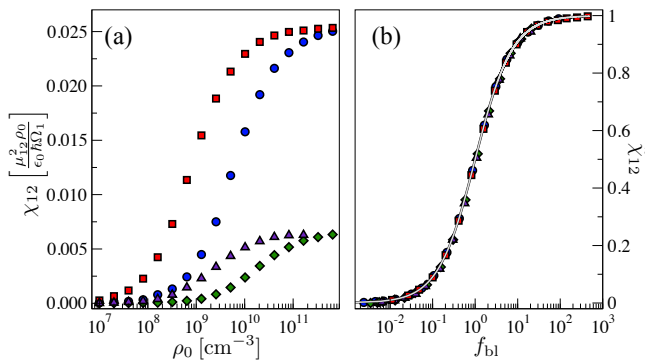}
\caption{(a) Nonlinear absorption coefficient ${\rm Im}[{\chi_{12}}]$ as a function of the density of Rubidium atoms for $\Omega_2=2$MHz and $\gamma_{12}=\gamma_{23}=0$.  The different symbols correspond to different principal quantum numbers and probe Rabi frequencies: $\Omega_1=1$MHz, $n=50$ (squares), $\Omega_1=0.5$MHz, $n=50$ (circles), $\Omega_1=1$MHz, $n=70$ (triangles) and $\Omega_1=1$MHz, $n=50$ (diamonds). (b) After proper scaling [cf. eqs.(\ref{scale_dens}) and (\ref{scale_chi})] all data points follow the simple universal scaling relation eq.(\ref{universal}) (line).}
 \label{chi_universal}
\end{center}
\end{figure}

It thus appears reasonable to employ the derived high-density limit as a natural scale for $\chi_{12}$. In addition we can use the interaction blockade of Rydberg excitation \cite{lfc01} to rescale the gas density, i.e. the abscissa in Fig.\ref{fig3}. Comparing the Rydberg atom density $\rho_{\rm ryd}=\rho_0N^{-1}\sum_i\rho_{33}^{(i)}$ obtained from the Monte Carlo calculation to the corresponding value $\rho_{\rm ryd}^{(0)}$ for vanishing interactions one obtains the fraction
\begin{equation}\label{scale_dens}
f_{\rm bl}=\frac{\rho_{\rm ryd}^{(0)}}{\rho_{\rm ryd}}-1
\end{equation}
of suppressed Rydberg excitations. Re-expressing the total atomic density through eq.(\ref{scale_dens}) and scaling the probe beam absorption by eq.(\ref{scale_chi}), yields a universal dependence of $\tilde{\chi}_{12}=\chi_{12}/\chi_{12}^{(\infty)}$ on the laser parameters, atomic density and interaction strength. We have verified this behavior for a wide range of parameters, some of which are exemplified in Fig.\ref{chi_universal} for $\gamma_{12}=\gamma_{23}=0$ (Fig.\ref{chi_universal}a). Indeed, all data points collapse on a single curve (Fig.\ref{chi_universal}b) described by
\begin{equation}\label{universal}
\tilde{\chi}_{12}=\frac{f_{\rm bl}}{1+f_{\rm bl}}\;.
\end{equation}
This simple formula nicely illustrates the effects of excitation blocking on the optical susceptibility of the Rydberg-EIT medium: For $f_{\rm bl}<1$ one finds a nonlinear absorption proportional to the blockade fraction, which, however, saturates at the two-level limit $\chi_{12}^{(\infty)}$ for $f_{\rm bl}\gtrsim1$, i.e. at the onset of a strong excitation blockade. Since $f_{\rm bl}\sim\Omega_1^4$ for weak Rydberg excitation \cite{hnp10} this implies a finite cubic nonlinearity, as observed in \cite{pgw10}.

In conclusion, we have presented a numerical method that permits to explore the optical response of an EIT medium in the presence of strong atomic interactions. In particular, we considered, Rydberg-Rydberg atom interactions, which yield strong photon-photon interactions, manifested in a nonlinear $\Omega_1$-dependence of the optical susceptibility. At small atomic densities our Monte Carlo results are in excellent agreement with reduced-density matrix calculations, but in addition are shown to be applicable over a much wider range of densities and interaction strengths. Beyond the present calculations, the described approach permits efficient simulations of very large atom numbers, which in future studies should enable detailed comparisons with experiments. Besides illuminating the effect of Rydberg interactions, the revealed universal behavior of the probe absorption may be of use in experiments, by permitting quick estimates of optical nonlinearities and allowing to determine the medium absorption from pure population measurements.
While we have primarily focussed on nonlinear absorption phenomena the developed approach can also be used to study the optical response of far-detuned EIT media and, thus, to explore prospects for realizing strong coherent photon-photon interactions.


\begin{thebibliography}{99}
\bibitem{EIT_rev} M. Fleischhauer, A. Imamoglu, and J.P. Marangos, Rev. Mod. Phys. {\bf 77}, 633 (2005).
\bibitem{hafi99} S. E. Harris, J. E. Field, and A. Imamoglu, Phys Rev. Lett. {\bf 64}, 1107 (1990).
\bibitem{luim00} M. D. Lukin and A. Imamoglu, Phys. Rev. Lett. {\bf 84}, 1419 (2000).
\bibitem{bhb09} M. Bajcsy et al., Phys. Rev. Lett. {\bf 102}, 203902 (2009).
\bibitem{moja07} A.K. Mohapatra, T.R. Jackson, and C.S. Adams, Phys. Rev. Lett. {\bf 98}, 113003 (2007).
\bibitem{moba08} A.K. Mohapatra et al., Nature Phys. {\bf 4}, 890 (2008).
\bibitem{wepr08} K.J. Weatherill et al., J. Phys. B {\bf 41}, 201002 (2008).
\bibitem{rahe09} U. Raitzsch et al., New J. Phys. {\bf 11}, 055014 (2009).
\bibitem{pgw10} J.D.Pritchard et al., Phys. Rev. Lett. 105, 193603 (2010)
\bibitem{fde10} R. Fleischhaker, T.N. Dey, and J. Evers, Phys. Rev. A {\bf 82}, 013815 (2010).
\bibitem{sgh10} H. Schempp et al., Phys. Rev. Lett. {\bf 104}, 173602 (2010).
\bibitem{welo08} H. Weimer, et al., Phys. Rev. Lett. {\bf 101}, 250601 (2008).
\bibitem{pode10} T. Pohl, E. Demler, and M.D. Lukin, Phys. Rev. Lett. {\bf 104}, 043002 (2010).
\bibitem{slm10} J. Schachenmayer, I. Lesanovsky, A. Micheli and A.J. Daley, New J. Phys. {\bf 12} 103044 (2010).
\bibitem{Rydexp} D. Tong et al., Phys. Rev. Lett. {\bf 93}, 063001 (2004); K. Singer et al., Phys. Rev. Lett. {\bf 93}, 163001 (2004); T. Vogt et al., Phys. Rev. Lett. {\bf 97}, 083003 (2006); R.  Heidemann et al., Phys. Rev. Lett. 99, 163601 (2007).
\bibitem{lfc01} M.D. Lukin et al., Phys. Rev. Lett. {\bf 87}, 037901 (2001).
\bibitem{mmm08} D. M\/oller, L.B. Madsen, and K. M\/olmer, Phys. Rev. Lett. {\bf 100}, 170504 (2008).
\bibitem{mlw09} M. M\"uller et al., Phys. Rev. Lett. {\bf 102}, 170502 (2009).
\bibitem{ae75} L. Allen, and J.H. Eberly, {\it Optical Resonance and Two-level Atoms} (Wiley, New York, 1975).
\bibitem{wila60} L.R. Wilcox and W.E. Lamb, Phys. Rev. {\bf 119}, 1915 (1960).
\bibitem{app07} C. Ates et al., Phys. Rev. A {\bf 76}, 013413 (2007).
\bibitem{hnp10} N. Henkel, R. Nath, and T. Pohl Phys. Rev. Lett. {\bf 104}, 195302 (2010).
\end{thebibliography}
\end{document}